\documentclass[aps,prb,preprint]{revtex4}

\usepackage{graphicx}
\usepackage{graphics}
\usepackage{epsfig}

\begin{document}
\title{ Transport equation for 2D electron liquid under microwave radiation
plus magnetic field and the Zero Resistance State }

\author{Tai-Kai Ng and Lixin Dai}
\address{Department of Physics, Hong Kong University of Science and
Technology, Clear Water Bay Road, Hong Kong}

\begin{abstract}

  A general transport equation for the center of mass motion is constructed to study transports of electronic system
  under uniform magnetic field and microwave radiation. The equation is applied to study 2D electron system in the
  limit of weak disorder where negative resistance instability is observed when the radiation field is strong
  enough. A solution of the transport equation with spontaneous AC current is proposed to explain the experimentally
  observed Radiation-Induced Zero Resistance State.

  \end{abstract}


\maketitle

\narrowtext

   The discovery of the zero-resistance states (ZRS) in two dimensional electron gas under uniform magnetic
 field\cite{mani,zudov} and microwave radiation has triggered a lots of theoretical\cite{t1,t2,t3,t4,t5,t6,t7} and
 experimental activities\cite{e1,e2} to understand the origin of this nontrivial state. Most of the theoretical work
 suggested that the origin of the ZRS is closely related to a negative-resistance instability that occurs in the
 system due to the combined effect of quantized Landau levels and photon-assisted scattering\cite{t1,t2,t3,t4,t7}.
 It was proposed that the ZRS can be explained if the current-dependent resistance of the system which becomes
 negative at small current (for strong enough microwave radiation) becomes positive again when the current
 $\vec{j}$ becomes large enough\cite{t1,t2,t3,t4}. The above physics was put together phenomenologically into an
 equation
 \begin{equation}
 \label{pheno}
  \vec{E}_d=\rho_H[\vec{j}\times\hat{z}]+R[|\vec{j}|]\vec{j}
  \end{equation}
 where $R[j]$ is a phenomenological current-dependent resistance which is negative at $j=0$, increases as a function
 of $j$ and passes through zero at $|\vec{j}|=j_o$\cite{t4}, $\vec{E}_d$ is the applied DC electric field and
 $\rho_H$ is the ordinary Hall resistivity. It was shown that equation \ (\ref{pheno}) admits
 time-independent, stripe-like spatially inhomogeneous solution which leads to zero differential resistance
 for net DC current less than a threshold value\cite{t4}. An obvious theoretical question is whether
 Eq.\ (\ref{pheno}) with the required property of $R[j]$ can be derived microscopically. This is the
 subject of this paper.

  Starting from first principle we shall derive in the following a transport equation for the center of mass velocity
 $\vec{v}=\vec{j}/ne$ that treats the effect of radiation to all order with the only expansion parameter in the
 problem being the strength of disorder. We note that a transport equation can also be derived from a Quantum
 Boltzmann Equation approach\cite{t7}. However it is difficult to obtain clear, analytical result in this approach
 because of the intrinsic complexity of the Boltzmann equation formulation itself. We shall see that the equation of
 motion for the center of mass offers a much simpler alternative.

  Our approach to the transport equation begins from the known observation that there exists an exact, one-to-one
 mapping between the solution of the Schr$\ddot{o}$dinger equation of a (charged) many-particle system in the absence
 of the microwave radiation and in the presence of the radiation for a class of Hamiltonian\cite{t2,t5,t6,t7,t8},
 \begin{equation}
 \label{h1}
 H(t)=H_o-e\sum_{i}\vec{r}_i.\vec{E}(t)
 \end{equation}
 where
 \begin{equation}
 \label{h2}
 H_o(\vec{r}_i)=\sum_{i}{1\over2m}(\vec{p}_i-{e\over c}\vec{A}(\vec{r}_i))^2+{1\over2}\sum_{i\neq
 j}V(\vec{r}_i-\vec{r}_j)+{1\over2}\sum_{i}\vec{r}_i.\tensor{K}.\vec{r}_{i}
 \end{equation}
 where $\vec{p}_i=-i\bar{h}\nabla_i$ is the canonical momentum for the $i^{th}$ particle in the system.
 $\vec{A}(\vec{r})=-{1\over2}(\vec{r}\times\vec{B})$ is the vector potential corresponding to a uniform, time-dependent
 magnetic field, $\vec{E}(t)$ is a time-dependent uniform electric field and $V(\vec{r})$ is the interaction
 potential between particles. The last term represents an external harmonic potential acting on the
 particles.

   The physics of the exact mapping can be seen by performing a coordinate transformation to
 the center of mass frame of the many-particle system. In the non-relativistic limit,
 the wavefunctions in the laboratory and CM frames are related by
 \[
 \psi_{lab}(\vec{r}_i;t)=\psi_{CM}(\vec{r}_i-\vec{R}(t);t)e^{i\theta(\vec{R}(t))},
 \]
 where $\vec{R}(t)={1\over N}\sum_{i}<\vec{r}_i(t)>$ is the center of mass coordinate in the laboratory
 frame, $N=$ number of particles and $\theta(\vec{R}(t))$ is an overall phase that depends on $\vec{R}(t)$ only.
 The corresponding Hamiltonian in the CM frame is
 \begin{eqnarray}
 \label{hcm}
 H'(t) & = & \sum_{i}{1\over2m}(\vec{p}_i'-m\dot{\vec{R}}(t)-{e\over c}\vec{A}_{CM}'(\vec{r}_i'))^2+{1\over2}\sum_{i\neq
 j}V(\vec{r}_i'-\vec{r}_j')  \\  \nonumber
  & & +{1\over2}\sum_{i}(\vec{r}_i'+\vec{R}(t)).\tensor{K}.(\vec{r}_i'+\vec{R}(t))-e\sum_{i}\vec{r}_i'.\vec{E}'(t)
 \end{eqnarray}
 where $\vec{r}'=\vec{r}-\vec{R}(t)$, $\vec{p}_i'=-i\bar{h}\nabla'$ and $\vec{A}'(\vec{r}')=-{1\over2}\vec{r}'
 \times\vec{B}'$. $\vec{B}'=\vec{B}$ and $\vec{E}'(t)=\vec{E}(t)+{1\over c}\dot{\vec{R}}(t)\times\vec{B}$ are the
 Galilean transformed magnetic and electric fields in the CM frame valid in the non-relativistic limit.
 The Schr$\ddot{o}$dinger equation in the CM frame, $i\hbar\dot{\psi}_{CM}=H'(t)\psi_{CM}$ can be simplified by a
 performing a gauge transformation
 $\psi_{CM}=\phi_{CM}e^{i\Lambda(t)}$, where $\hbar\Lambda(t)=\sum_{i}\left(m\dot{\vec{R}}(t).\vec{r}'-{1\over2}
 \int^tdt'\vec{R}(t').\tensor{K}\vec{R}(t')\right)$. With Eq.\ (\ref{hcm}), we obtain
 \begin{equation}
 i\hbar{\partial\over\partial t}\phi_{CM}(\vec{r}';t)=\left(H_o(\vec{r}'_i)-\sum_i\vec{r}'_i.\vec{a}(t)\right)
 \phi_{CM}(\vec{r}';t),
 \end{equation}
 where $\vec{a}(t)=e\left(\vec{E}(t)+{1\over c}(\dot{\vec{R}}(t)\times\vec{B})\right)-\tensor{K}.\vec{R}(t)
 -m\ddot{\vec{R}}(t)=0$. $\vec{a}(t)$ vanishes because it is the equation of motion for $\vec{R}(t)$ derived
 directly from the corresponding Heisenberg equation of motion. We thus arrive at the conclusion
 that for the class of Hamiltonian \ (\ref{h1}), there exist a one-to-one mapping between the solution of
 the Schr$\ddot{o}$dinger equation in the presence of the microwave radiation $\psi(\vec{r}_i;t)$, and in the
 absence of the radiation $\phi_{CM}(\vec{r}_i';t)$, where $\psi(\vec{r}_i;t)=\phi_{CM}(\vec{r}_i-\vec{R}(t);t)
 e^{i\Lambda(t)}$. Physically, for the particular form of Hamiltonian we considered, the wavefunction of the system
 follows the center of mass motion rigidity in the presence of the radiation field.

  The above result suggests that for more general Hamiltonians of form $H_G(t)=H(t)+U$, a new perturbation scheme
 where the microwave radiation is treated exactly to all order can be set up by treating $U$ as perturbation. The
 perturbation scheme can be set up most easily in the center of mass frame. We shall consider static impurity
 potential $U=\lambda\sum_{i}u(\vec{r}_i)$ in the following. Notice that a static potential become time-dependent
 in the CM frame and should be treated by time-dependent perturbation theory.

   To derive the transport equation we start from the exact Heisenberg equation of motion for the center of mass
 coordinate, $i\hbar\dot{\vec{R}}(t)=[\vec{R}(t),H_G(t)]$ with $\tensor{K}=0$\cite{t2}. We obtain after some simple
 algebra
 \begin{equation}
 \label{em1}
 m\ddot{\vec{R}}(t)=e\left(\vec{E}(t)+{1\over c}
 (\dot{\vec{R}}(t)\times\vec{B})\right)-{\lambda\over N}\int d^dr\left(\nabla u(\vec{r})\right)n(\vec{r},t)
 \end{equation}
 where $n(\vec{r},t)=\sum_i<\delta(\vec{r}-\vec{r}_i)>$ is the time-dependent average electron density. In the CM
 frame where $n(\vec{r},t)=n_{CM}(\vec{r}-\vec{R}(t),t)$, we obtain for small $\lambda$ from linear-response theory
 \[
 n_{CM}(\vec{r}',t)=n^{(0)}_{CM}(\vec{r}')+\lambda\int
 d^dr"dt"\chi(\vec{r}'-\vec{r}";t-t')u(\vec{r}"+\vec{R}(t')), \]
 where $\chi(\vec{r}'-\vec{r}";t-t')$ is the (equilibrium) retarded density-density response function in CM frame
 derived with the Hamiltonian $H_o$. Going back to the laboratory frame and performing the disorder-average
 $<u(\vec{r}>=0$ and $<u(\vec{r})u(\vec{r}')>=|u|^2\delta(\vec{r}-\vec{r}')$, we obtain to second order in $\lambda$,
 an impurity-averaged equation of motion for $\vec{R}(t)$ in laboratory frame\cite{t2},
\begin{equation}
 \label{em1}
 m\ddot{\vec{R}}(t)=e\left(\vec{E}(t)+{1\over c}
 (\dot{\vec{R}}(t)\times\vec{B})\right)+\alpha\nabla_{\vec{R}(t)}\int dt'\chi(\vec{R}(t)-\vec{R}(t');t-t')
 \end{equation}
 where $\bar{n}=N/V$ and $\alpha=\lambda^2|u|^2/\bar{n}$. The equation is manifestly gauge invariant and suggests
 that to second order in the impurity potential, the effects of particle statistics and interaction are reflected
 only in the density-density response function. In the following we shall apply this equation to study the ZRS in
 2D electron systems.

  In this case we consider $\vec{E}(t)=\vec{E}_o\sin(\omega t)+\vec{E}_d$, where $\vec{E}_d$
 is a small DC electric field. To simplify the equation we divide the center of mass motion into "fast" and
 "slow" parts,
 \begin{equation}
 \label{trialsol}
 \vec{R}(t)=\vec{R}_f(t)+\vec{R}_s(t)\sim\vec{A}_o\cos(\omega t)+\vec{B}_o\sin(\omega t)+\vec{R}_s(t).
 \end{equation}
 $\vec{R}_f(t)$ describes the center of mass motion induced by the radiation field whereas $\vec{R}_s(T)$ describes
 induced motion under the DC field $\vec{E}_d$. The two kinds of motions are coupled by the impurity scattering
 term which is a non-linear function of $\vec{R}(t)$. We shall assume that the coupling between $\vec{R}_s(t)$
 and $\vec{R}_f(t)$ does not modify qualitatively the behavior of $\vec{R}_f(t)$ and the main effect of coupling
 is to produce an effective equation of motion for $\vec{R}_s(t)$. Notice that we keep only the first harmonic
 terms in $\vec{R}_f(t)$ in Eq.\ (\ref{trialsol}). This is valid in the small
 $\vec{E}_o$ limit when the size of the orbit $R_c\sim\sqrt{\vec{A}_o^2+\vec{B}_o^2}$ is much less than the magnetic
 length $l=(\hbar c/eB)^{1/2}$ and is justifiable under the experimental condition\cite{mani,zudov} where the
 magnetic field is weak and the magnetic length is very long.

    To treat the impurity scattering term we write $\nabla_{\vec{R}(t)}\int_{-\infty}^t
  dt'\chi(\vec{R}(t)-\vec{R}(t');t-t')= -\int \vec{q}d^dq\int {d\omega'\over\pi}Im\chi(\vec{q},\omega')
  \int_{-\infty}^tdt'e^{i[\vec{q}.(\vec{R}(t)-\vec{R}(t'))-\omega'(t-t')]}$,
  where $\chi(\vec{q},\omega)$ is the Fourier transform of $\chi(\vec{r};t)$. To evaluate the integral over $t'$
  we make the local approximation $\vec{R}_s(t)-\vec{R}_s(t')\sim \vec{v}(t)(t-t')+{1\over2}\dot{\vec{v}}(t)
  (t-t')^2$, and
  \[
   \int_{-\infty}^tdt'e^{i[\vec{q}.(\vec{R}(t)-\vec{R}(t'))-\omega'(t-t')]}\sim
   \int_{-\infty}^tdt'e^{i[\vec{q}.(\vec{R}_f(t)-\vec{R}_f(t'))+\left(\vec{q}.\vec{v}(t)-\omega'\right)(t-t')]}
   \left(1+i\vec{q}.\dot{\vec{v}}(t){(t-t')^2\over2}\right)   \]
  valid for $\vec{v}(t+T')-\vec{v}(t)<<\vec{v}(t)$, where $T'\sim2\pi/\omega_B$ is the characteristic time-scale
  for the density response. The $\vec{R}_f(t)-\vec{R}_f(t')$
  term can be expanded in a series of $\vec{A}_o, \vec{B}_o$ using the identity
  $e^{ix\sin\theta}=\sum_mJ_m(x)e^{im\theta}$. We obtain after some algebra,
 \begin{eqnarray}
 \label{force}
 \nabla_{\vec{R}(t)}\int_{-\infty}^t dt'\chi(\vec{R}(t)-\vec{R}(t');t-t') & = &
  \sum_{m,m'}e^{i(m-m')(\omega t+\theta)}\int\vec{q}d^dqJ_m(z(\vec{q}))J_{m'}(z(\vec{q}))
   \\ \nonumber
  & &
  \times\left(1-i{\vec{q}.\dot{\vec{v}}(t)\over2}{\partial^2\over\partial\omega'^2}\right)
  \chi(\vec{q},\omega')_{\omega'=m\omega+\vec{q}.\vec{v}(t)}
 \end{eqnarray}
  where $z(\vec{q})=\sqrt{(\vec{q}.\vec{A}_o)^2+(\vec{q}.\vec{B}_o)^2}$ and $\tan\theta=\vec{q}.\vec{A}_o/
  \vec{q}.\vec{B}_o$\cite{t2}. We shall consider $z(\vec{q})^2=R_c^2q^2/2$ in the following corresponding to
  circularly polarized light\cite{t7}. In this case the response of the system is isotropic and the computation
  is much simplified. The effective force on $\vec{R}_s(t)$ is derived from the $m=m'$ terms in \ (\ref{force}).
  To order $(R_c/l)^2$ we keep only $m(m')=0,1$ terms. Putting it back into the equation of motion \ (\ref{em1}),
  we obtain
  \begin{equation}
  \label{treq}
  \left(m+\Pi[\omega,|\vec{v}(t)|]\right)\dot{\vec{v}}(t)=e\vec{E}_d+{e\over c}(\vec{v}(t)\times\vec{B})
  -R[\omega,|\vec{v}(t)|]\vec{v}(t),
  \end{equation}
  where $\omega$ is the frequency of the microwave radiation, and
  \begin{eqnarray}
  \label{math3}
  R[\omega,\vec{v}] & \sim & {\alpha\over v^2}\int\vec{q}.\vec{v}d^dq\left[(1-{z(\vec{q})^2\over2})Im\chi(\vec{q},\vec{q}.\vec{v})
  +{z(\vec{q})^2\over4}Im\chi(\vec{q},\omega+\vec{q}.\vec{v})\right],
  \\  \nonumber
  \Pi[\omega,\vec{v}] & \sim & -{\alpha\over4}\int q^2d^dq{\partial^2\over\partial\omega'^2}\left[(1-{z(\vec{q})^2\over2})
   Re\chi(\vec{q},\omega')_{\omega'=\vec{q}.\vec{v}}+{z(\vec{q})^2\over4}
   Re\chi(\vec{q},\omega')_{\omega'=\omega+\vec{q}.\vec{v}}\right].
  \end{eqnarray}

   It is obvious that $R[\omega,\vec{v}]$ represents dissipative response of the electron gas to external
 perturbations whereas $\Pi[\omega,\vec{v}]$ is an effective mass correction on the center of mass motion coming
 from the corresponding reactive response. Negative contributions to the resistance shows up in the second term of
 $R$ for density-density response functions with resonant structure, $\chi(\vec{q},\omega)\sim\sum_n
 {g_n(\vec{q})\over\pi}{-2\omega_n\over\omega^2-\omega_n^2}$, where $\omega_n>0$ are resonant energy levels and
 $g_n(\vec{q})$ is a positive definite function. In this case, $R\sim\sum_{n}\left(R_{1n}\omega_n+R_{2n}
 (\omega_n-\omega)\right)$, where $R_{1n}$ and $R_{2n}$ are positive definite numbers. We see that negative
 contributions to $R_{2}$ exist for $\omega_n<\omega$. The physical origin of the negative
 resistance has been discussed in several earlier works\cite{t1,t2,t3,t4,t7} and we shall not repeat them here.
 Correspondingly, the effective mass contribution from level $n$ is positive (negative) when $\omega>(<)\omega_n$.
 We note that the effective mass correction originates from the impurity scattering term is of order
 $\Pi[\omega,v]\sim m\times(1/\omega_B\tau)<<m$ in the weak-disorder limit, where $\omega_B=eM/mc$ is the cyclotron
 frequency and $\tau\sim(g_{2d}\lambda^2|u|^2)^{-1}$ is the elastic lifetime. Eq.\ (\ref{treq}) differs from the
 phenomenological equation \ (\ref{pheno}) mainly in the presence of the inertial term $m\dot{\vec{v}}$ which allows
 the system to admit time-dependent solutions in the present case.

  To see whether $R[\omega,v]$ has the expected behavior we consider density response function of non-interacting
  gases where\cite{t2} $\chi(\vec{q},\omega)=(2\pi l^2)^{-1}\sum_{m,n}(n_2!/n_1!)\left(Q^2/2\right)^{n_1-n_2}
  e^{-(Q^2/2)}\left(L^{n_1-n_2}_{n_2}(Q^2/2)\right)^2\left({n_F(n\omega_B)-n_F(m\omega_B)\over
 \omega+(n-m)\omega_B+i\delta}\right)$,
 where $n_1=\max(n,m)$, $n_2=\min(n,m)$, $L^n_m(x)$ is the associated Laguerre polynomials, $n_F(\epsilon)$ is the
 Fermi distribution function and $Q^2=|\vec{q}|^2l^2$. To incorporate inelastic lifetime effects we also introduce
 a phenomenological broadening $\Gamma(T)$ to the Landau levels, i.e.
 $\delta(\epsilon-n\omega_B)\rightarrow(\pi)^{-1}\Gamma/((\epsilon-n\omega_B)^2+\Gamma^2)$.
 $R[\omega,\vec{v}]$ is evaluated numerically with these approximations. In figure one we present numerical
 results for the normalized resistance $R(\omega,v)/R(0,0)$, as a function of normalized velocity
 $v_N=v/(l\omega_B)$ for $\omega/\omega_B=0,0.85,1.0,1.1,1.2$, taking $(R_c/l)^2=0.1$, $E_F\sim 10\omega_B$,
 $T=2\omega_B$, $\Gamma=0.5$ and keeping $20$ levels in the sum. We observe that $R(\omega,v\rightarrow0)$ becomes
 negative when $\omega\gtrsim\omega_B$, in agreement with previous results\cite{t7}. For
 $\omega>\omega_B$ $R(\omega,v)$'s increases and cross zero at around $v_{oN}\sim0.05$. The effect of microwave
 radiation decreases rapidly for $v_N\gtrsim0.2$. These qualitative behaviors of
 $R(\omega,v)$ are in agreement with expectation and are not modified by changing $T$ or $\Gamma$.
 \begin{figure}
 \includegraphics[width=6cm, height=6cm, angle=-90]{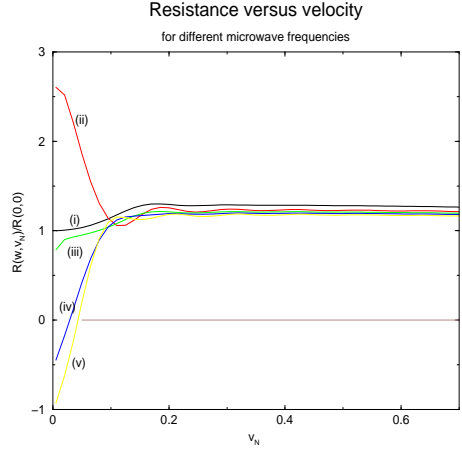}
 \caption{\label{Fig.1} Normalized resistance as a function of velocity for $\omega/\omega_B=0(i),0.85(ii),
 1.0(iii),1.1(iv),1.2(v)$.}
 \end{figure}

   Eq.\ (\ref{treq}) allows time-dependent solutions. In the absence of the DC field, a simple, spatially homogeneous
  solution which allows the system to stabilize itself around the point $v_o=(l\omega_B)v_{oN}$ is
 \[
 \vec{v}^{(0)}(t)=R_o(\cos(\omega_st)\hat(x)+\sin(\omega_st)\hat(y)), \]
 with $R_o=v_o/\omega_s$ where $\omega_s=eB/(m+\Pi(\omega,v_o))c\sim\omega_B$. The solution represents a collective
 circular motion of the whole fluid moving with speed $v_o$. Notice that a time-independent, spatially inhomogeneous
 solution corresponding to a pattern of alternating current stripes\cite{t6} still exist. However this solution is
 energetically less favorable because it requires a higher energy to create the charge inhomogeneity
 needed to maintain the stripes of currents.

  To understand the ZRS we have to consider the boundaries of the droplet of electron fluid. For sharp boundaries
 the boundary condition $j_{\bot}=0$ has to be imposed where $j_{\bot}$ is the component of current perpendicular to
 the boundary. As a result an edge region with a time-independent current $j_{//}\sim j_o$ must form. The size of
 this region is determined by the microscopic charge dynamics\cite{t4} which is still undetermined. Nevertheless
 according to Eq.\ (\ref{treq}) an electric field perpendicular to the boundary with magnitude $\sim E_d=Bv_o/c$ has
 to be present in this region to maintain the steady current flow. A similar edge region also exists at the opposite
 edge with a current running in the opposite direction, rather similar to edge states in Quantum Hall Effect.

   A state with a small net current flow can be created with minimal disturbance to the system by shrinking the size
 of one edge region and enlarging the other. In this case, the net voltage drop across the sample is given by
 \[
 V_y=\int E(y)dy={B\over c}\int v_x(y)dy={B\over ne^2c}\int j_x(y)dy=\rho_HI_x,
 \]
 corresponding to a resistance matrix with $\rho_{xx}=0$ and $\rho_{xy}=\rho_H$, i.e. the ZRS.

  Some comments about the validity of our theory is in order. We note that in deriving Eq.\ (\ref{treq}) we made the
 local approximation which assumes slowly varying $\vec{v}(t)$ whereas the spontaneous current state we propose
 oscillate with a frequency $\sim\omega_B$. The oscillatory solution is allowed because of the presence of the inertia
 term $m\dot{\vec{v}}(t)$. Our analysis shows that the correction to the inertia term is small
 ($\sim m/(\omega_B\tau)$) in the limit of weak disorder and the local approximation mainly affects $R[\omega,v]$.
 Therefore our general description of the ZRS should remain valid as long as the qualitative property of
 $R[\omega,v]$ is correct. Another simplification we employed in our analysis is the assumption of circularly
 polarized light. The response of the system which is isotropic in this approximation would become anisotropic when
 this assumption is relaxed\cite{t2,t7}.

   Lastly we made a comment on the macroscopic nature of the spontaneous current state we proposed. We note that
 in general a spontaneous current state with $|\vec{v}(t)|=v_o$ is characterized by a position and time dependent
 (2D) unit vector field $\hat{n}(\vec{x},t)$ representing the direction of the current. The order parameter
 field $\hat{n}(\vec{x},t)$ has the same symmetry as the ordinary 2D $x-y$ model, or superfluids. The main
 difference between the ZRS state and superfluids is that the rigidity of the order parameter is protected by
 repulsive interaction in the case of superfluids, whereas it is protected by the principle of least dissipation in
 the ZRS. The similarity between the two systems suggests that the two systems may share some common macroscopic
 features. For example, vortex-like solitonic excitations may exist in the ZRS and may lead to the residue
 resistance $\sim R_oexp(-(T_o/T))$ observed in the ZRS state\cite{mani,zudov}. The existence and nature of solitonic
 excitations depends on the detailed current dynamics of the ZRS state and will be investigated in a coming paper.

 {\it Acknowledgements}
 This work is supported by HKRGC through Grant 602803.


\references
 \bibitem{mani} R. Mani {\em et.al.}, Nature {\bf 420}, 646 (2002).
 \bibitem{zudov} M.A. Zudov, R.R. Du, L.N. Pfeiffer and K.W. West, \prl {\bf 90}, 0468071(2003).
 \bibitem{t1} A.C. Durst, S. Sachdev, N. Read and S.M. Girvin, \prl {\bf 91}, 086803(2003).
 \bibitem{t2} X.L. Lei and S.Y. Liu, \prl {\bf 91}, 226805(2003).
 \bibitem{t3} Junren Shi and X.C. Xie, \prl {\bf 91}, 086801(2003).
 \bibitem{t4} A.V. Andreev, I.L. Aleiner and A.J. Millis, \prl {\bf 91}, 056803(2003).
 \bibitem{t5} K. Park, \prb {\bf 69}, 201301(2004).
 \bibitem{t6} J. I$\tilde{n}$arrea and G. Platero, \prl {\bf 94}, 016806(2005).
 \bibitem{e1} C.L. Yang {\em et.al.}, \prl {\bf 91}, 096803(2003).
 \bibitem{e2} J. Zhang {\em et.al.}, \prl {bf 92}, 156802(2004).
 \bibitem{t7} M.G. Vavilov and I.L. Aleiner, \prb {\bf 69}, 035303(2004);
 I.A. Dmitriev {\em et.al.}, \prb {\bf 71}, 115316(2005).
 \bibitem{t8} J.F. Dobson, \prl {\bf 73}, 2244(1994).
 \end{document}